\documentclass[showpacs,amsmath,amssymb,twocolumn,floatfix,prl]{revtex4}
\usepackage[dvips]{graphicx}
\input{epsf}

\begin{document}

\title{Quantum computation of the Anderson transition in 
presence of imperfections}


\author{A. A. Pomeransky and D. L. Shepelyansky}
\homepage[]{http://www.quantware.ups-tlse.fr}
\affiliation{Laboratoire de Physique Th\'eorique, UMR 5152 du CNRS, 
Univ. Paul Sabatier, 31062 Toulouse Cedex 4, France}
\date{June 30, 2003}

\begin{abstract}
We propose a quantum algorithm for simulation of 
the Anderson transition in disordered lattices
and study numerically its sensitivity to 
static imperfections in a quantum computer. 
In the vicinity of the critical point the algorithm
gives a quadratic speedup 
in computation of diffusion rate and localization length,
comparing to the known classical algorithms.
We show that the Anderson transition can be detected
on  quantum computers with $7 - 10$ qubits.

\end{abstract}

\maketitle

The problem of metal-insulator transition 
of noninteracting electrons in a disordered potential
was pioneered by Anderson in 1958 \cite{anderson58}. 
Since then  it continues to attract an active interest of 
researchers all over the world (see {\it e.g.} \cite{lee85,kramer,mirlin}
and Refs. therein). 
In addition to analytical and experimental studies
of the problem an important contribution to the understanding of 
its properties was made with the help of numerical simulations based on
various computational methods adapted to the physics of this phenomenon.
Indeed, the numerical studies allowed to obtain some values of
critical exponents in the vicinity of the transition 
and to study certain system characteristics at the critical
point including level spacing statistics and conductance fluctuations
for the cases of different symmetries and  system dimensions 
(see {\it e.g.} 
\cite{kramer,mirlin,shklovski,zharekeshev,slevin99}).
These numerical simulations are performed with the help
of modern supercomputers and are at the border of their 
computational capacity.

The recent  progress in quantum computation demonstrated that 
due to quantum parallelism certain tasks can be performed
much faster on a quantum computer (see \cite{chuang} and Refs. therein).
The most known example is the Shor algorithm for factorization of 
large integers \cite{shor} which is exponentially faster than any 
known classical algorithm.  A number of efficient quantum algorithms 
was also proposed for simulation of quantum evolution
of certain Hamiltonians  including many-body quantum systems
\cite{lloyd,ortiz} and problems of quantum chaos
\cite{schack,georgeot,benenti}. In Ref. \cite{georgeot}
it was shown that the evolution propagator
in a regime of dynamical or Anderson localization
can be simulated efficiently on a quantum computer.
However, the algorithm proposed there requires a significant
number of redundant qubits and is not accessible  for an experimental
implementation with a first generation of quantum computers 
composed of 5 - 10 qubits. 

In this paper we propose a quantum algorithm for a
quantum dynamics in the regime of
Anderson localization. This algorithm requires no redundant qubits
thus using the available $n_q$ qubits in an optimal way.
The propagation on a unit time step is performed 
in $O(n_q^2)$ quantum elementary gates while any known classical algorithm
requires $O(2^{n_q})$ operations 
for a vector of size $N=2^{n_q}$. Due to these properties
the Anderson transition can be already detected
on a quantum computer with 7 - 10 qubits.
The basic elements of the algorithm involve one qubit rotations,
controlled phase shift $C(\phi)$, controlled-NOT gate $C_{N}$ and 
the Quantum Fourier Transform (QFT)  \cite{chuang}. 
All these quantum operations
have been already realized for 3 - 7 qubits 
in the NMR-based quantum computations reported in Refs. \cite{cory,chuang1}.
Thus the main obstacle for experimental
detection of the Anderson transition in quantum computations
is related to the effects of external decoherence \cite{paz} and
residual static imperfections \cite{georgeot1}
which restrict the number of available quantum gates.
The results obtained for operating quantum algorithms
\cite{benenti,terraneo} show that the effects of 
static imperfections affect the accuracy of quantum computation
in a stronger way comparing to the case of random noisy gate errors.
Due to that in this paper we concentrate our studies on the
case of static imperfections investigating their impact on the
system properties in the vicinity of the Anderson transition.

To study the effects of static imperfections in quantum computations
of the Anderson transition we choose the generalized kicked rotator model 
described by the unitary evolution of the wave function $\psi$:
\begin{equation}
\begin{array}{c}
\bar{\psi}={\hat U} \psi= \exp (-i V(\theta,t)) 
\exp ( -i  H_{0}({\hat n})) \psi  \;\; .
\end{array}
\label{eq1}
\end{equation}
Here $\bar{\psi}$ is the new value of $\psi$ after one map
iteration given by the unitary operator ${\hat U}$,
$H_{0}(n)$ gives the rotational phases in the basis of
momentum ${\hat n} = -i \partial/\partial \theta$,
the kick potential $V(\theta,t)$ depends
on the rotator phase $\theta$  and time $t$ 
measured in number of kicks,  $\psi(\theta+2\pi) = \psi(\theta)$.
For $V(\theta,t) = k \cos \theta$ and $H_0 = Tn^2/2$
one has the kicked rotator model described in detail in \cite{izrailev}.
The evolution given by (\ref{eq1}) results from the Hamiltonian
$H=H_0(n)+V(\theta,t)\delta_1(t)$, where $\delta_1(t)$ is a periodic 
$\delta$-function with period 1 and $(n,\theta)$ are conjugated variables.
In the case when  the potential $V(\theta,t) = 
-2\tan^{-1} (2k (\cos\theta + \cos {\omega_1 t} + \cos {\omega_1 t}))$ 
depends quasi-periodically on time $t$ the model can be exactly reduced to
the three-dimensional (3D) Lloyd model \cite{3f}. Indeed, the time dependence
of $V(\theta,t)$ can be eliminated by introduction of
extended phase-space  with a replacement
$H_{0} \rightarrow H_0(n) + \omega_1 n_1 +\omega_2 n_2$.
Then the  linear dependence on quantum numbers $n_{1,2}$
gives fixed frequency rotations of the conjugated phases
$\theta_{1,2}=\omega_{1,2}t$. The extensive studies performed
in \cite{3f} showed that this model displays the Anderson metal-insulator
transition at $k = k_c \approx 0.5$ with the critical exponents being
close to the values found in other 3D solid state models. In this paper
following \cite{borgonovi} we choose in (\ref{eq1}) the potential
$V(\theta,t) = k(1+ 0.75 \cos\omega_1 t \cos\omega_2t) \cos \theta$
with $\omega_1=2\pi\lambda^{-1}$, $\omega_2=2\pi\lambda^{-2}$ and
$\lambda=1.3247...$ being the real root of the cubic equation
$x^3-x-1=0$. The rotation phases $H_0(n)$ are randomly 
distributed in the interval $(0,2\pi)$. 
This model shows the Anderson transition
at $k_c \approx 1.8$ \cite{borgonovi} 
with the characteristics similar to those
of the Lloyd model studied in \cite{3f}.

\begin{figure}[t!]  
\centerline{\epsfxsize=4.0cm\epsffile{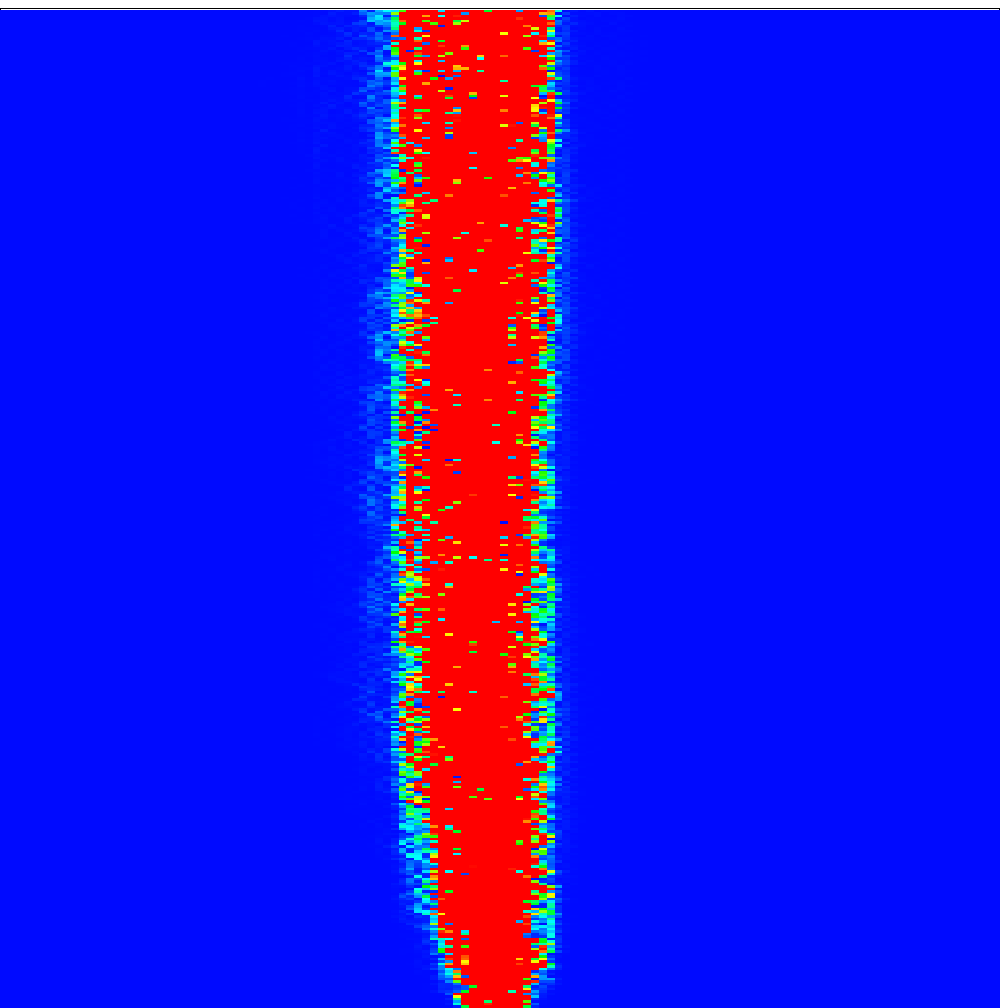}
\hfill\epsfxsize=4.0cm\epsffile{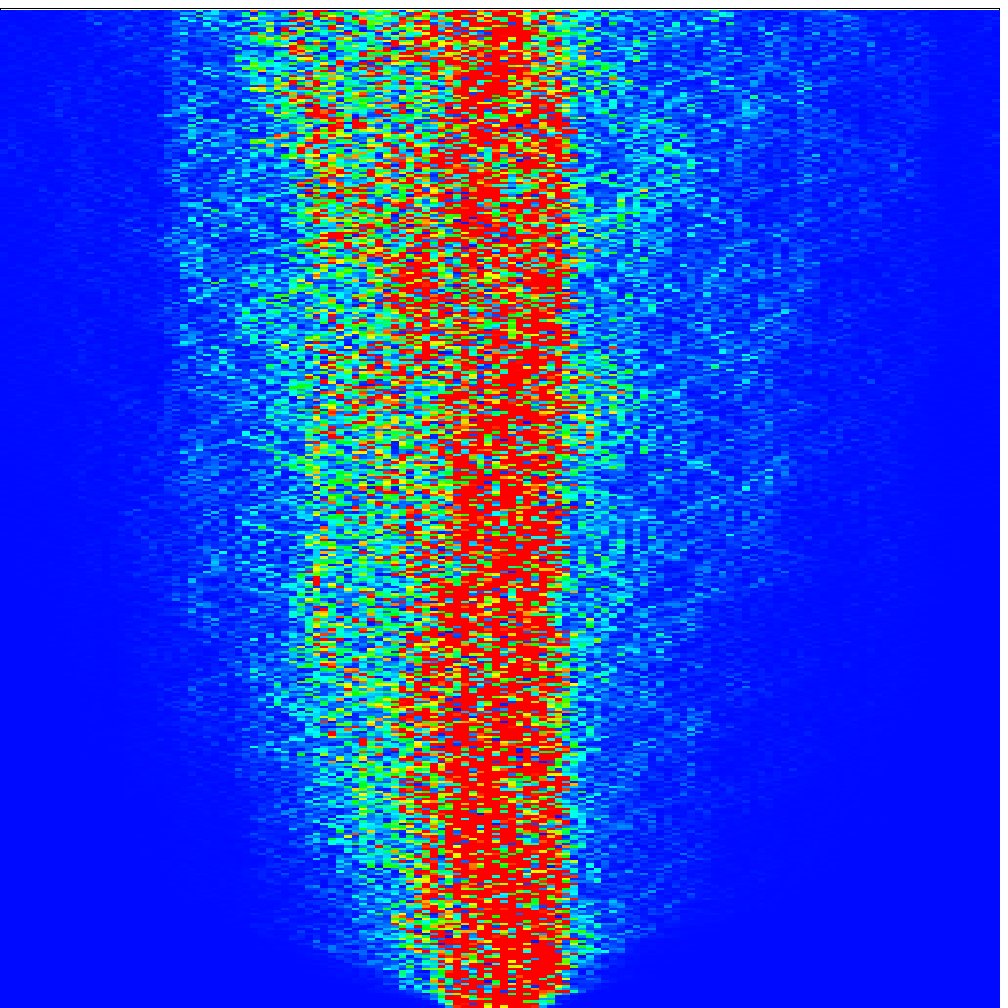}}
\centerline{\epsfxsize=4.0cm\epsffile{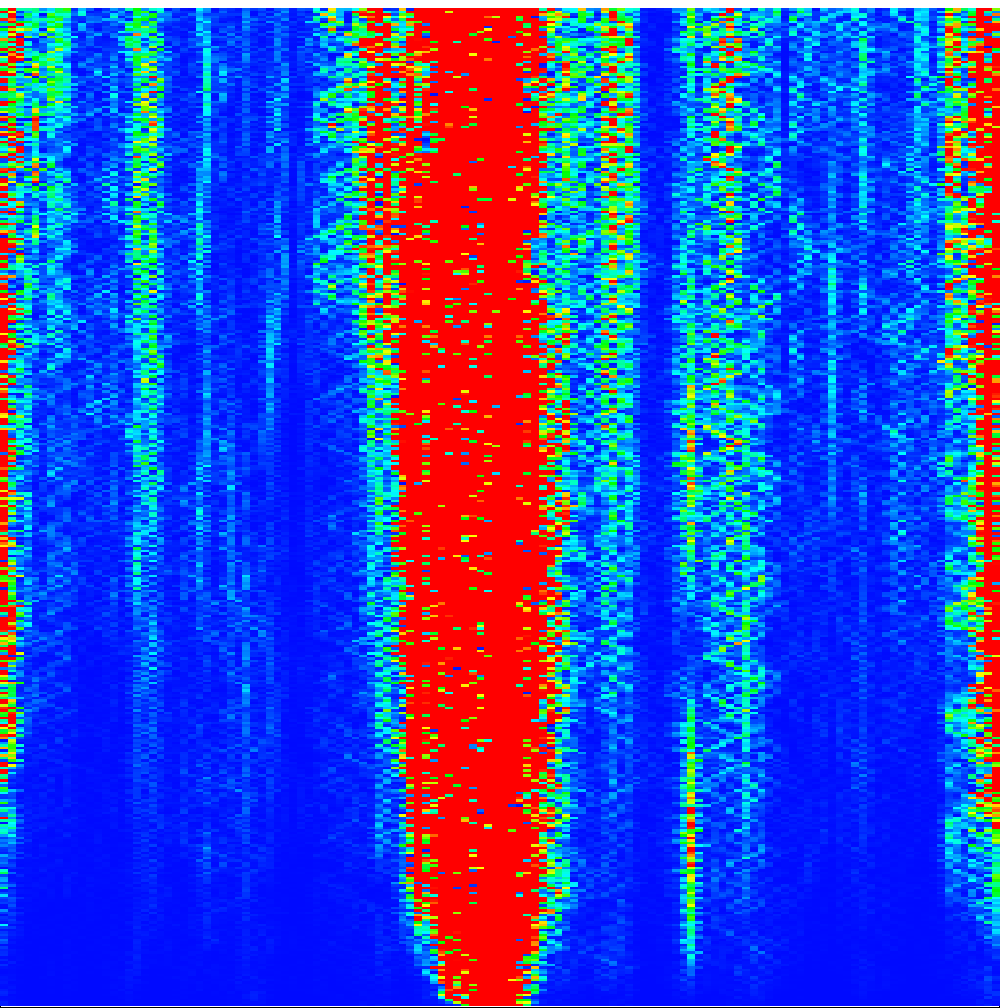}
\hfill\epsfxsize=4.0cm\epsffile{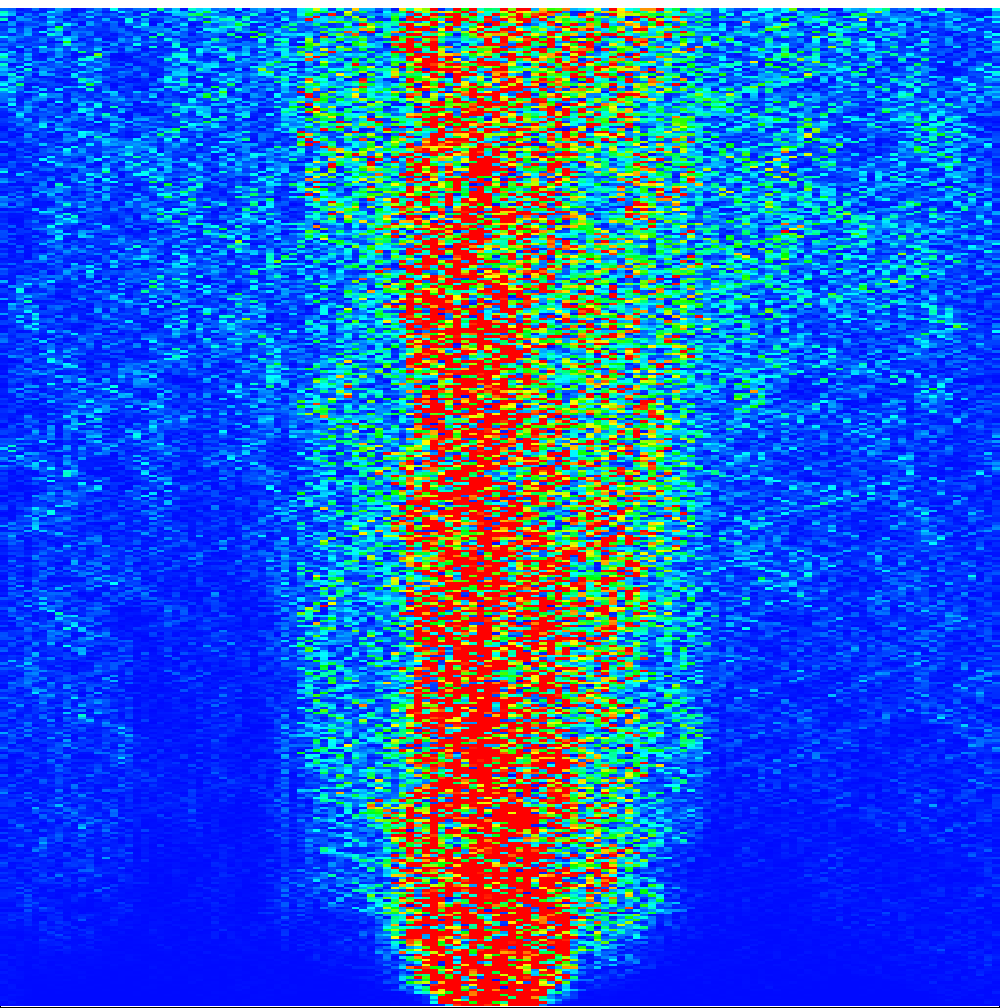}}
\vglue -0.3cm 
\caption{(color online) 
The time evolution of the probability distribution
$|\psi_n|^2$ in the localized (left column, $k=1.2$) and delocalized
(right column, $k=2.4$) phases for $n_q=7$ qubits ($N=2^{n_q}$),
with $0\leq t \leq 400$ (vertical axis) and 
$-N/2 < n \leq N/2$ (horizontal axis); $k_c = 1.8$. The color is
proportional to probability: blue/black for zero  and
red/white for maximal values. The strength of static imperfections 
is $\epsilon=\mu=0$ for top row and 
$\epsilon=\mu=10^{-4}$ for bottom row. 
}
\label{fig1}       
\end{figure}

The quantum algorithm simulating the time evolution of this model 
is constructed in the following way. The quantum states $n=0,...,N-1$
are represented by one quantum register with $n_q$ qubits so that 
$N=2^{n_q}$.  The initial state with all probability at $n_0=0$
corresponds to the state $|00...0>$ 
(momentum $n$ changes on a circle with $N$ levels).
The phase  rotation $U_T=\exp(-iH_0(n))$ in the momentum
basis $n$ is performed with the help of quantum random phase
generator built from two unitary operators $U^{(1)}_T$ and 
$U^{(2)}_T$. The operator 
$U^{(1)}_T =\prod_{j=1}^{n_q} e^{i \phi_j \sigma^z_j} $
gives rotation of qubit $j$ by a random phase $\phi_j$.
Here and below $\sigma^x, \sigma^y, \sigma^z$ are Pauli matrices.
To improve the independence of quantum phases we then apply
the operator $U^{(2)}_T = \prod_{k=1}^{M} C_N(i_{M-k},j_{M-k}) 
\prod_{k=1}^{M} e^{i\phi_{j_k}^{\prime}\sigma^z_{j_k}}C_N(i_k,j_k)$.
This transformation represents a random sequence with $M$
one-qubit phase shifts $e^{i\phi_{j_k}^{\prime}\sigma^z_{j_k}}$
and controlled-NOT gates
$C_N(i_k,j_k)$
followed by the inversed sequence of controlled-NOT
gates $C_N(i_{M-k},j_{M-k})$. Here $C_N(i_k,j_k)$ inverts the 
qubit $j_k$ if  the qubit $i_k$ is 1; $i_k, j_k $ and 
phases $\phi_{j_k}^{\prime}$ are chosen randomly.
The resulting random quantum phase
generator $U_T= U_T^{(2)}  U_T^{(1)}$ gives more and more
independent random phases with the increase of $M$.
We use $M \approx 2 n_q$ (at $n_q \approx 10$) that according to our tests
generates good random phase values. This step involves
$3M+n_q$ quantum gates. 
After that the kick operator 
$U_k= \exp(-ik(t) \cos \theta)$ is performed as follows.
First, with the help of the QFT the wave function is 
transformed from momentum $n$ to phase $\theta$ representation
in $O(n_q^2/2)$ gates.  Then $\theta$ can be written in the binary
representation as $\theta/2\pi = 0.a_1a_2..a_{n_q}$ with 
$a_i=0$ or 1. It's convenient to use the notation
$\theta=\pi a_1 + \bar{\theta}$ to single out the most significant
qubit. Then due to the relation
$\cos\theta=(-1)^{a_1}\cos\bar{\theta}=\sigma^z_{1}\cos\bar{\theta}$
the kick operator takes the form 
$U_k=e^{-ik(t)\cos\theta} = e^{-i\sigma^z_{1} k(t)\cos\bar{\theta}}$,
where $\sigma^{(z,x)}_{1}$ act on the first qubit.
This operator can be approximated to an arbitrary precision
by a sequence of one-qubit gates applied to the first qubit
and the diagonal operators $S^m= e^{i m a_1\bar{\theta}}$.
The $S-$operators are given by the product of $n_q-1$ two-qubit
gates
as $S^m=\prod_{j=2}^{n_q} C_{1,j}(\pi m 2^{-j+1})$
where controlled phase shift  gate $C_{j_1,j_2}(\phi)$
makes a phase shift $e^{i\phi}$ if both qubits $j_{1,2}$ are 1.
Then we introduce the unitary operator
$R_{\gamma}(\bar{\theta})=H S^{1} H\; e^{-i\frac{\gamma}{2}\sigma^z_1} \; 
H S^{-2} H \; e^{-i\frac{\gamma}{2}\sigma^z_1} \; H S^{1} H$
where $H=(\sigma^z_1+\sigma^x_1)/\sqrt{2}$ is the Hadamard gate.
It can be exactly reduced to the form 
$R_{\gamma}(\bar{\theta})= \cos^2\frac{\gamma}{2}-
\sin^2\frac{\gamma}{2}\cos(2\bar{\theta})-
i\sigma^z_1 \sin\gamma \cos(\bar{\theta}) 
+i\sigma^x_1\sin^2\frac{\gamma}{2}\sin(2\bar{\theta})$
and hence for small $\gamma$ we have
$R_{\gamma}(\bar{\theta}) = e^{-i\sigma^z_1\gamma\cos\bar{\theta}}+
i\sigma^x_1\frac{\gamma^2}{4}\sin(2\bar{\theta})+O(\gamma^3)$.
The term with $\gamma^2$ can be eliminated using the symmetric representation
$R_{\gamma/2}(\bar{\theta})R_{\gamma/2}(-\bar{\theta}) =
H S^{1} H\; e^{-i\frac{\gamma}{4}\sigma^z_1} \; 
H S^{-2} H \; e^{-i\frac{\gamma}{2}\sigma^z_1} \; H S^{2} H \;
e^{-i\frac{\gamma}{4}\sigma^z_1} \; H S^{-1} H
=e^{-i\sigma^z_1\gamma\cos(\bar{\theta})}+O(\gamma^3)$.
Thus the kick operator is given by
$U_k=(R_{\gamma/2}(\bar{\theta})R_{\gamma/2}(-\bar{\theta}))^l+O(l \gamma^3)$
where the number of steps $l = k/\gamma$ and we used 
in our numerical simulations the small
parameter $\gamma =k/l \approx 0.2$ that gives $l \approx 5 -  10$
for $k \sim 1 - 2$. After that the state is transfered to the momentum 
representation by the QFT. Thus an iteration
(\ref{eq1}) is performed for 
$2^{n_q}$ states in $n_g$ elementary gates where
$n_g = 2[k/\gamma](n_q+2)+n_q^2+6n_q+3M+9$ with the square brackets
denoting the integer part. This algorithm is optimal for the kicked rotator
model with moderate values of $k$ where 
$n_g$ value remains reasonable. It can be easily generalized to $d$ dimensions.

In our numerical simulations we study the effects of static
quantum computer imperfections considered in \cite{georgeot1,benenti,terraneo}.
In this case all gates are perfect
but between gates $\psi$ accumulates a
phase factor  $e^{i {\hat \varphi} }$
with ${\hat \varphi} = \sum_j (\eta_j \sigma^z_j+
\mu_j\sigma^x_j \sigma^x_{j+1})$. Here $\eta_j, \mu_j$ vary randomly with
$j=1,...,n_q$, $\eta_j$ represents static  one-qubit energy shifts,
$-\epsilon/2 \le \eta_j  \le \epsilon/2$,
and $\mu_j$ represents static inter-qubit couplings on a circular chain,
$-\mu/2 \le \mu_j  \le \mu/2$.

\begin{figure} 
\epsfxsize=3.0in
\epsffile{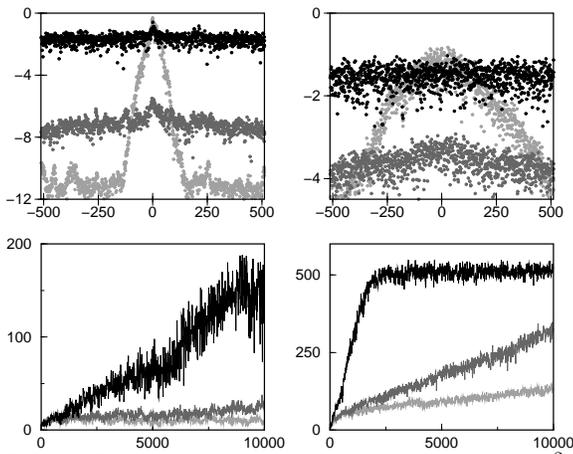}
\vglue -0.4cm
\caption{Top row: logarithm of probability $\log_{10} |\psi_n|^2$ vs. 
momentum $n$ after $t=10000$ iterations; dark gray  curves are shifted
down by 5 (left) and 2 (right). Bottom row: dependence of IPR $\xi$ on time $t$.
The left/right  column corresponds to localized/delocalized phase 
at $k=1.2$ and $k=2.4$ respectively. 
The three curves represent
$\epsilon=0; 2\times 10^{-5}; 6\times 10^{-5}$
with color changing from light gray to black with increase of $\epsilon$;
$\mu=\epsilon$, $n_q=10$.
}
\label{fig2}       
\end{figure} 
\vglue -0.0cm

An example of time evolution of probability distribution
in the momentum representation $n$ is shown in Fig.~1.
Below the Anderson transition  ($k<k_c$)
the probability remains bounded near initial value $n_0$,
while above it ($k>k_c$) a diffusive spreading in $n$ takes place.
Comparing to the ideal quantum computation the static imperfections
lead to probability transfer on levels located far away from
the center of the wave packet. This effect is related to the structure
of the QFT where a mismatch in the quantum gates generates
high harmonics.  As a result static imperfections create
a plateau in the probability distribution which
level grows with the increase of $\epsilon$ and $\mu$
(see Fig.~2). This leads to an artificial diffusion
of the second moment of the distribution 
$<n^2>=<\psi_n|(n-n_0)^2|\psi_n>$. Since the plateau in probability
extends over all $N$ levels the rate of this diffusion grows
exponentially with $n_q$ at fixed $\epsilon, \mu$ (data not shown). A similar 
effect was discussed in \cite{song} for the quantum computation
of the kicked rotator with noisy gates. Due to that the most
appropriate characteristic to study is the Inverse Participation Ratio (IPR)
$\xi$ which is extensively used in systems 
with localization \cite{kramer,mirlin}
and which determines the number of levels on which 
the wave function is concentrated $(\sum_n |\psi_n|^4 = 1/\xi)$.
In contrast to $<n^2>$, the IPR $\xi$ remains stable with respect to
noise in the gates during polynomially large times \cite{song}.

\begin{figure}[t!]
\epsfxsize=3.0in
\epsffile{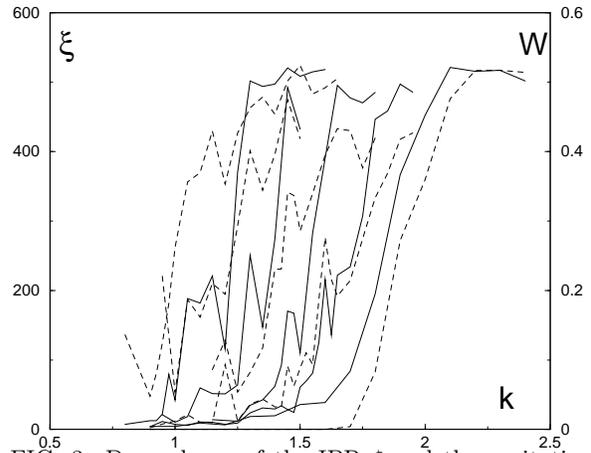}
\vglue -0.5cm
\caption{Dependence of the IPR $\xi$ and the excitation probability $W$
(full and dashed curves for left and right scales respectively)
on the kick strength $k$
for $n_q=10$ and $t \geq 10^5$,
$\epsilon = 0; 10^{-5}; 2\times 10^{-5};
4\times 10^{-5}; 8\times 10^{-5}$
(corresponding to curves from right to left); $\mu=0$.
}
\label{fig3}       
\end{figure}

The variation of $\xi$ with time and $\epsilon, \mu$ is shown
in Fig.~2. For moderate imperfections,  during  a rather long time interval
$\xi$ remains close to its value in the exact algorithm.
However, at very large times $t \geq 10^5$ it saturates at some
value which depends on $k$ and $\epsilon, \mu$. A typical 
example of such a dependence is presented in Fig.~3.
Here, $\xi$ shows a sharp jump from small ($\xi \sim 1$)
to large ($\xi \sim N$) values which takes place
in a narrow interval of $k$ values. This is a manifestation of the Anderson
transition from localized to delocalized states. 
The critical point $k_c$  can be numerically defined as a such value
of $k$ at which $\xi$ is at the middle between its two 
limiting values. The data of Fig.~3 show that 
the critical point $k_c(\epsilon)$
decreases with the increase of the strength of imperfections.
The physical origin of this effect is related to the additional
transitions induced by static imperfections which
naturally lead to a delocalization at a lower value of $k$
compared to the ideal computation.  Another method to detect the
position of the critical point $k_c(\epsilon)$
in presence of imperfections is to measure the two most significant qubits
which code the value of momentum $n$. After a few tens
of measurements of first 2 qubits one determines the probability 
$W =\sum_{n=(N/4,3N/4)}|\psi_n|^2$. At sufficiently large $t$
this probability shows a sharp jump  from a value $W=0$
to $W \approx 0.5$ when $k$ is varied. This  allows
to determine the critical point 
and gives the values of $k_c(\epsilon)$ close to those 
obtained via IPR $\xi$ (see Fig.~3).

\begin{figure}[t!]
\epsfxsize=3.0in
\epsffile{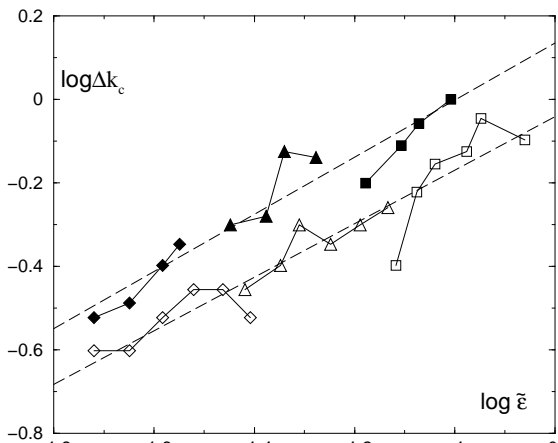}
\vglue -0.5cm
\caption{Dependence of the shift of the critical
point $\Delta k_c(\epsilon)=k_c-k_c(\epsilon)$ on rescaled imperfection
strength $\tilde{\epsilon} = \epsilon n_g \sqrt{n_q}$
for $\epsilon=2\times 10^{-5}$ (diamonds), 
$4\times 10^{-5}$ (triangles) and 
$8\times 10^{-5}$ (squares); open/full symbols are for
$\mu=0$, $8 \leq n_q \leq 13$ and 
$\mu=\epsilon$, $8 \leq n_q \leq 11$ respectively;
$k_c=1.8$. The dashed lines show the scaling relation (\ref{eq2}).
}
\label{fig4}       
\end{figure}

The shift of the critical point 
$\Delta k_c(\epsilon) = k_c-k_c(\epsilon)$ depends on 
$\epsilon, \mu$ and $n_q$. From the IPR data obtained
for various $\epsilon, \mu, n_q$, see Fig.~4, we find that
the global parameter dependence can be described 
by the scaling relation 
\begin{equation}
\Delta k_c(\epsilon) = A \tilde{\epsilon }^{\; \alpha}, \;\; 
{\tilde \epsilon} = \epsilon n_g \sqrt{n_q}
\label{eq2}
\end{equation}
The data fit gives $A=3.0$, $\alpha = 0.64$
for $\mu =0$ and $A=4.8$, $\alpha = 0.68$ for $\mu = \epsilon$.
This result can be understood from the following arguments.
According to \cite{benenti,terraneo} the time scale
$t_f$, on which the fidelity of quantum computation is close to 
unity,  is determined by the parameter $\tilde{\epsilon}$ 
($t_f \sim 1/\tilde{\epsilon}$).  Thus, an effective matrix
element induced by static imperfections
between ideal localized eigenstates can be estimated as
$U_{ef} \sim \tilde{\epsilon} Q \sim \tilde{\epsilon} /l^{\beta}$,
where $Q$ is a typical  overlap of localized eigenstates
which for the Anderson localization in $d$ dimensions can be 
estimated as $Q \sim l^{-\beta}$ with $\beta=d/2$  and
$l$ being the localization length for the exact algorithm (see a
discussion in \cite{benenti1} for $d=1$). 
The imperfections induced delocalization should take place
when $U_{ef}$ exceeds the level spacing in a block 
of size $l$ ($U_{ef} > \Delta_l \sim 1/l^d$).
Taking into account that near the critical point 
the localization length scales as 
$l \sim \Delta k^{-\nu}$ with $\nu \sim 1.5$ 
(see \cite{kramer,3f,borgonovi}) we obtain that
$\alpha = 1/(\nu (d - \beta)) = 2/\nu d $.
The obtained value of $\alpha $ would give a reasonable value
of $\nu \approx 1.0$ but in our model (\ref{eq1}) the situation 
is more complicated.
Indeed, the dynamics in (\ref{eq1}) takes place in one dimension
and hence one expects $\beta=1/2$ and $\nu \approx 0.6$.
The later value has a noticeable difference from 
a usually expected value \cite{kramer,3f,borgonovi}.
A possible reason for this discrepancy can be related
to the fact that in the algorithm the perturbations
give far away transitions (see Fig.~1) which effectively decrease
the value of $\beta$, also near the critical point
the correlations in the matrix 
elements can play an important role. Further studies are required to
clarify this point.

Finally, let us note that in the vicinity of  critical point 
the number of states grows with time as 
$n^d \sim t$ \cite{kramer,mirlin,slevin99}. Hence, the number of classical
operations for $t$ kicks
can be estimated as $n_{gcl} \sim t N^d \log^d N \sim 
t^2 \log^d t$ while the quantum algorithm will need
$n_g \sim d  n_q^2 t \sim t \log^2 t$ gates assuming $d$ quantum registers
with $N^d=2^{d n_q} \sim t$ states. 
The coarse-grained characteristics of the probability distribution
can be determined from few measurements of most significant
qubits, {\it e.g.} $W$ as in Fig.3.
Thus, even if each step in (\ref{eq1})
is efficient, the speedup is only quadratic
near the critical point. Above the critical point we have diffusive
growth with $n^d \sim t^{d/2}$ and the speedup is stronger:
$n_{gcl} \sim n_g^{(1+d/2)}$ for $d>2$.

This work was supported in part by the NSA and ARDA under ARO 
contract No. DAAD19-01-1-0553 and the EC IST-FET project EDIQIP.
We thank CalMiP at Toulouse and  IDRIS at Orsay for access 
to their supercomputers.

\end{document}